\documentclass[3p,times,procedia]{elsarticle}
\usepackage{nupha_ecrc}

\volume{00}

\firstpage{1}

\journalname{Nuclear Physics A}

\runauth{}

\jid{nupha}

\jnltitlelogo{Nuclear Physics A}

\usepackage{amssymb,amsmath,bm,latexsym,amsfonts}

\usepackage[figuresright]{rotating}

\newcommand{\beq}{\begin{eqnarray}}
\newcommand{\eeq}{\end{eqnarray}}

\begin{document}

\begin{frontmatter}



\dochead{}

\title{Analytic approaches to relativistic hydrodynamics}


\author{Yoshitaka Hatta}

\address{Yukawa Institute for Theoretical Physics, Kyoto University, Kyoto 606-8502, Japan}

\begin{abstract}
I summarize our recent work towards finding and utilizing analytic solutions of relativistic hydrodynamic. In the first part I discuss various exact solutions of the second-order conformal hydrodynamics. In the second part I compute flow harmonics $v_n$ analytically using the anisotropically deformed Gubser flow and discuss its dependence on $n$, $p_T$, viscosity, the chemical potential and the charge. 
\end{abstract}


\end{frontmatter}


\section{Introducution}
\label{}

{\it Why analytic hydro?} The past decade has witnessed a tremendous success of relativistic hydrodynamics in describing observables of heavy-ion collisions \cite{deSouza:2015ena}. Nowadays, a number of sophisticated numerical codes for solving the hydrodynamic equation exist. Together with the realistic initial condition and the QCD equation of state, they can fit the bulk of heavy-ion data at RHIC and the LHC quite well. In such circumstances, it is easy to get an impression that there is not much one can do analytically. 

Yet, there are multiple reasons to study analytic solutions of the hydrodynamic equation. Firstly, they provide physical intuition into the problem. There are famous solutions such as the Hubble flow for the expansion of the universe and the Bjorken flow for the expansion of fireballs in heavy-ion collisions. These solutions, while different from reality in details, are something one always keeps in mind as the zeroth approximation.  Secondly, the hydrodynamic equation is an interesting and fascinating subject in its own right from a mathematical viewpoint. Many analytic solutions of the ideal and viscous hydrodynamic equations have been found over a century. Yet, a complete understanding of the Navier-Stokes equation remains one of the most challenging problems of modern mathematics. 
 Thirdly, there are interesting questions which numerical approaches cannot fully answer. For example, `How do flow harmonics $v_n$ {\it functionally} depend on $n$, or viscosity?' It would be interesting if there is a kind of `pocket formula' for the $n$-dependence of $v_n$. Last but not least, analytic solutions are useful for testing the accuracy of numerical codes, especially for viscous hydrodynamics. 
  
In this presentation, I summarize our recent work towards finding and utilizing analytic solutions of relativistic hydrodynamics \cite{Hatta:2014gqa,Hatta:2014gga,Hatta:2014jva,Pang:2014ipa,Hatta:2015kia,Hatta:2015era,Hatta:2015hca}.  The main goal is to demonstrate that there are actually a lot of things one can do analytically.  In the first part, I  construct exact solutions of the second-order conformal hydrodynamic equation. In the second part, I compute flow harmonics $v_n$ analytically for the anisotropically deformed Gubser flow \cite{Gubser:2010ze,Gubser:2010ui}.   Some of the results have direct phenomenological implications and are worth pursuing in more elaborate numerical studies.


\section{Second-order hydrodynamics}

The hydrodynamic equation is the continuity equation for the energy momentum tensor
\beq
\nabla_\mu T^{\mu\nu}=0\,, \qquad T^{\mu\nu}=\varepsilon u^\mu u^\nu+p(g^{\mu\nu}+u^\mu u^\nu) + \pi^{\mu\nu}\,. \label{eq}
\eeq
$\pi^{\mu\nu}$ is the shear stress tensor relevant to viscous hydrodynamics. In the Navier-Stokes (first order) approximation, it is simply $\pi^{\mu\nu}=-2\eta \sigma^{\mu\nu}$ where $\eta$ is the shear viscosity. In the second order approximation, the precise form of $\pi^{\mu\nu}$ is still under active debate, but it typically contains a lot of terms. If one assumes conformal symmetry, the number of terms is reduced \cite{Baier:2007ix}. But its most general form is still very complicated
\beq
\pi^{\mu\nu}&=&-2\eta \sigma^{\mu\nu} + \tau_\pi\left(\Delta^\mu_\alpha \Delta^\nu_\beta D\pi^{\alpha\beta} + \frac{4}{3}\vartheta \pi^{\mu\nu}\right) +\lambda_2\pi^{\langle \mu}_{\ \lambda} \Omega^{\nu\rangle \lambda} \nonumber \\
&& \qquad + \lambda_1\pi^{\langle \mu}_{\ \lambda}\pi^{\nu\rangle \lambda}+\tau_\sigma \left(\Delta^\mu_\alpha \Delta^\nu_\beta D\sigma^{\alpha\beta}+\frac{1}{3}\sigma^{\mu\nu}\vartheta\right)
 -\tilde{\eta}_3 \sigma^{\langle \mu}_{\ \lambda}\sigma^{\nu\rangle \lambda}
-\tau_{\pi\pi}\sigma^{\langle \mu}_{\ \lambda}\pi^{\nu\rangle \lambda}  +\lambda_3 \Omega^{\langle \mu}_{\ \lambda}\Omega^{\nu\rangle \lambda}\,, \label{pi}
\eeq
where $\Omega^{\mu\nu}$ is the vorticity tensor and $\vartheta=\nabla_\mu u^\mu$ is the expansion. The Israel-Stewart equation corresponds to keeping only the first line.  
In the second line one may argue that $\pi^{\mu\nu}$ and $-2\eta \sigma^{\mu\nu}$ can be identified. However, this is valid only in the asymptotic Navier-Stokes regime which is not assumed here.

First, I will be interested in finding exact solutions of (\ref{eq}) together with (\ref{pi}).    
In general, finding analytic solutions of (\ref{eq}) is very difficult even in the ideal case $\pi^{\mu\nu}=0$. If $\pi^{\mu\nu}$ is given by (\ref{pi}) with all the transport coefficients assumed to be nonvanishing, it seems impossible to make any analytical progress. However, there is a trick. To explain this let me review the Gubser flow 
 
 \section{Gubser flow}
 
 One usually solves (\ref{eq}) in the Cartesian coordinates. or in the `Rindler' coordinates
\beq
ds^2=-dt^2+dx_1^2+dx_2^2+dx_3^2\,.
\eeq
If there is boost-invariance, it is often convenient to work in the `Rindler' coordinates
\beq
ds^2=-d\tau^2+dx_\perp^2+x_\perp^2d\phi^2+\tau^2dy^2\,, \label{1}
\eeq
 where $\tau=\sqrt{t^2-x_3^2}$ is the proper time, $y=\frac{1}{2}\ln \frac{t+x_3}{t-x_3}$ is the spacetime rapidity and $x_\perp=\sqrt{x_1^2+x_2^2}$.    If there is conformal symmetry, one can combine the above coordinate transformation with the Weyl transform  of the metric $g_{\mu\nu}(x)\to \Lambda^2(x)\hat{g}_{\mu\nu}(\hat{x})$  and solve the hydrodynamic equation in the $\hat{x}^\mu$ coordinates. Gubser's idea was to choose $\Lambda^2=\tau^{2}$  \cite{Gubser:2010ze} so that
 \beq
 d\hat{s}^2=\frac{ds^2}{\tau^2}=\frac{-d\tau^2+dx_\perp^2+x_\perp^2d\phi^2}{\tau^2}+dy^2=-d\rho^2+\cosh^2\rho (d\Theta^2+\sin^2\Theta d\phi^2) + dy^2\,. \label{de}
 \eeq
The resulting metric is that of the three-dimensional de Sitter space $dS_3$ and a flat dimension for $y$. In the last equality, the $dS_3$ part is written in  the so-called global coordinates.
In the latter coordinates, Gubser considered the simplest form of the flow velocity $(\hat{u}^\rho,\hat{u}^\Theta,\hat{u}^\phi,\hat{u}^y)=(1,0,0,0)$. With this ansatz, the ideal hydrodynamic equation $\nabla_\mu \hat{T}^{\mu\nu}=0$ can be solved very easily. The solution is then transformed back to Minkowski space 
\beq
\varepsilon \propto \frac{1}{\tau^{4/3}}\frac{1}{(L^4+2(\tau^2+x_\perp^2)+(\tau^2-x_\perp^2)^2)^{4/3}}\,. \label{gub}
\eeq
The parameter $L$ can be interpreted as the transverse size of the colliding nuclei. 
One recognizes that the factor $1/\tau^{4/3}$ is identical to the Bjorken flow, but the solution also has a nontrivial dependence on $x_\perp$. It is a boost-invariant, radially expanding solution. Remarkably, Gubser also derived an exact solution of the Navier-Stokes equation where $\pi^{\mu\nu}=-2\eta \sigma^{\mu\nu}$.

\section{Exact solutions of second-order hydrodynamics}
\subsection{Conformal soliton flow}

The Weyl transform  is clearly a very powerful technique to construct nontrivial solutions of the hydrodynamic equations. One can consider different functions $\Lambda^2(x)$ and arrive at different solutions. This has been explored in \cite{Hatta:2014gqa,Hatta:2014gga}. Here I consider one such transformation. Instead of dividing by $\tau^2$ as in (\ref{de}), let me divide by $x_\perp^2$
\beq
d\hat{s}^2=\frac{-dt^2+dx_\perp^2+dz^2}{x_\perp^2}+d\phi^2=-\cosh^2 \bar{\rho} d\bar{\tau}^2+d\bar{\rho}^2+\sinh^2\bar{\rho} d\bar{\Theta}^2+d\phi^2\,. \label{ant}
\eeq
Now I am in $AdS_3\times S^1$, the three-dimensional anti-de Sitter space $AdS_3$ times the unit circle. In the second equality the $AdS_3$ part is again written in the global coordinates where $\bar{\tau}$ plays the role of time.  The simplest flow velocity in this space is the static one  $(\hat{u}^{\bar{\tau}},\hat{u}^{\bar{\rho}},\hat{u}^{\bar{\Theta}},\hat{u}^\phi)=(1/\cosh\bar{\rho},0,0,0)$. This is actually a known solution in Minkowski space called `conformal soliton flow' first found in \cite{Friess:2006kw} and rediscovered in \cite{Nagy:2009eq,Hatta:2014gqa} using different methods. The latter works also found an exact rotating solution characterized by the flow velocity 
\beq
(\hat{u}^{\bar{\tau}},\hat{u}^{\bar{\rho}},\hat{u}^{\bar{\Theta}},\hat{u}^\phi)
=\left(\frac{1}{\sqrt{\cosh^2\bar{\rho}-\omega^2}},0,0,\frac{\omega}{\sqrt{\cosh^2\bar{\rho}-\omega^2}}\right)\,,
\eeq
where $\omega$ is the angular velocity. Naturally, this solution has nonvanishing $\Omega^{\mu\nu}$. 

So far the construction is entirely analogous to the Gubser flow. But the conformal solution flow is spherically symmetric and therefore has vanishing shear tensor $\sigma^{\mu\nu}=0$ even in the rotating case $\omega\neq 0$. It is then tempting to include the full second-order corrections (\ref{pi}) (of which many terms vanish) and try to find exact solutions. This is indeed possible.  In the non-rotating case, Refs.~\cite{Hatta:2014gqa,Hatta:2014gga} found the following solutions. 
 \beq
  \varepsilon_{2nd}&\propto& 
  \frac{1}{(L^2+(t+r)^2)^2(L^2+(t-r)^2)^2}
  \left(\frac{4L^2x_\perp^2}{(L^2+(t+r)^2)(L^2+(t-r)^2)}\right)^{\frac{9}{2(\lambda_1-3)}}\,, \nonumber
  \\
&\propto &\frac{1}{(L^2+(t+r)^2)^2(L^2+(t-r)^2)^2}
\left(1-\frac{4L^2x_\perp^2}{(L^2+(t+r)^2)(L^2+(t-r)^2)}\right)^{\frac{9}{2(\lambda_1-3)}}\,, \nonumber\\
&\propto &\frac{1}{(L^2+(t+r)^2)^2(L^2+(t-r)^2)^2}\left( \frac{4L^2x_\perp^2\bigl((L^2+(t+r)^2)(L^2+(t-r)^2)-4L^2x_\perp^2\bigr)}{(L^2+(t+r)^2)^2(L^2+(t-r)^2)^2}
\right)^{-\frac{9}{2(\lambda_1+6)}}\,, \nonumber \\
&\propto &\frac{1}{(L^2+(t+r)^2)^2(L^2+(t-r)^2)^2} \left(\frac{L^2r^2}{(L^2+(r+t)^2)(L^2+(r-t)^2)}\right)^{\frac{9}{2(\lambda_1-3)}}\left(\frac{r^2}{x_\perp^2}\right)^{\frac{9}{2(\lambda_1+6)}}\,, \nonumber \\
&\propto &\frac{1}{(L^2+(t+r)^2)^2(L^2+(t-r)^2)^2} \left(\frac{L^2r^2}{(L^2+(r+t)^2)(L^2+(r-t)^2)}\right)^{\frac{-9}{\lambda_1+6}}\,,
\label{viscous}
 \eeq
  where $r=|\vec{x}|$.  In the rotating case $\omega\neq 0$, the last two terms of (\ref{pi}) come into play and the equation is very hard to solve. 
  Nevertheless, exact solutions have been found in \cite{Hatta:2014gqa,Hatta:2014gga}. In terms of the energy density, it is given by  
  \beq
  \frac{\varepsilon_{2nd}}{\varepsilon_{ideal}} = \left|1+\frac{21b}{4}\right|^{\frac{2(105-32\lambda_1)}{7(4\lambda_1-21)}} \left|1+\lambda_1 b\right|^{\frac{18}{4\lambda_1-21}}\,,
  \eeq
 where $b(\bar{\rho})$ is essentially the inverse Reynolds number and is given by  the solution of the following implicit equation ($c$ is an integration constant)
 \beq
 b(21b+4)^{\frac{105-32\lambda_1}{7(4\lambda_1-21)}}(\lambda_1 b+1)^{1+\frac{9}{4\lambda_1-21}} =c \lambda_3\omega^2\frac{\sinh^2\bar{\rho}}{\cosh^2\bar{\rho}-\omega^2}\,.
 \eeq
Note that the solutions do not depend on $\tau_\pi$ and $\lambda_2$ in (\ref{pi}) because the corresponding terms vanish identically for these solutions. Other examples of exact solutions can be found in \cite{Hatta:2014gga}.

\subsection{Unorthodox Bjorken flow}

When $\sigma^{\mu\nu}$ is nonvanishing, the situation becomes much more complicated. As the simplest example, consider the Bjorken flow velocity 
$u^\mu=\delta^\mu_\tau$. This  has $\sigma^{\mu\nu}\neq 0$ and $\Omega^{\mu\nu}=0$, so most of the terms in (\ref{pi}) are relevant. While the equation looks quite daunting,  it turns out that there exists a deceptively simple exact boost-invariant solution. It is just a power-law
\beq
\varepsilon = \frac{C}{\tau^4}\,.
\eeq
The fourth power is unusual (cf., $\varepsilon \propto 1/\tau^{4/3}$ for the Bjorken flow), but it actually follows from the dimensional reason.  A more peculiar feature is that the normalization $C$ is not arbitrary, but is a unique number in a given theory. It is completely fixed by the transport coefficients due to the nonlinearity of the equation
\beq
C^{1/4}= \frac{3\eta-16\tau_\pi +2\tau_{\pi\pi} \pm \sqrt{ (3\eta-16\tau_\pi +2\tau_{\pi\pi})^2+4(4\lambda_1-3)(2\tau_\sigma+\tilde{\eta}_3)}}{4(4\lambda_1-3)}\,.\label{C4}
\eeq
(The transport coefficients are made dimensionless by factoring out an appropriate power of $\varepsilon$.) 

This solution is so exotic that one might wonder if it is an artifact of the complicated second-order formulation of  hydrodynamics. This is not so. Essentially the same solution ($\varepsilon\propto 1/\tau^4$ with a fixed normalization)  has been found  for the Boltzmann equation in the relaxation time approximation \cite{Hatta:2015kia}. The existence of this type of solution is also suggested by fluid-gravity correspondence. 

 \begin{figure}[bp]
   \includegraphics[width=70mm]{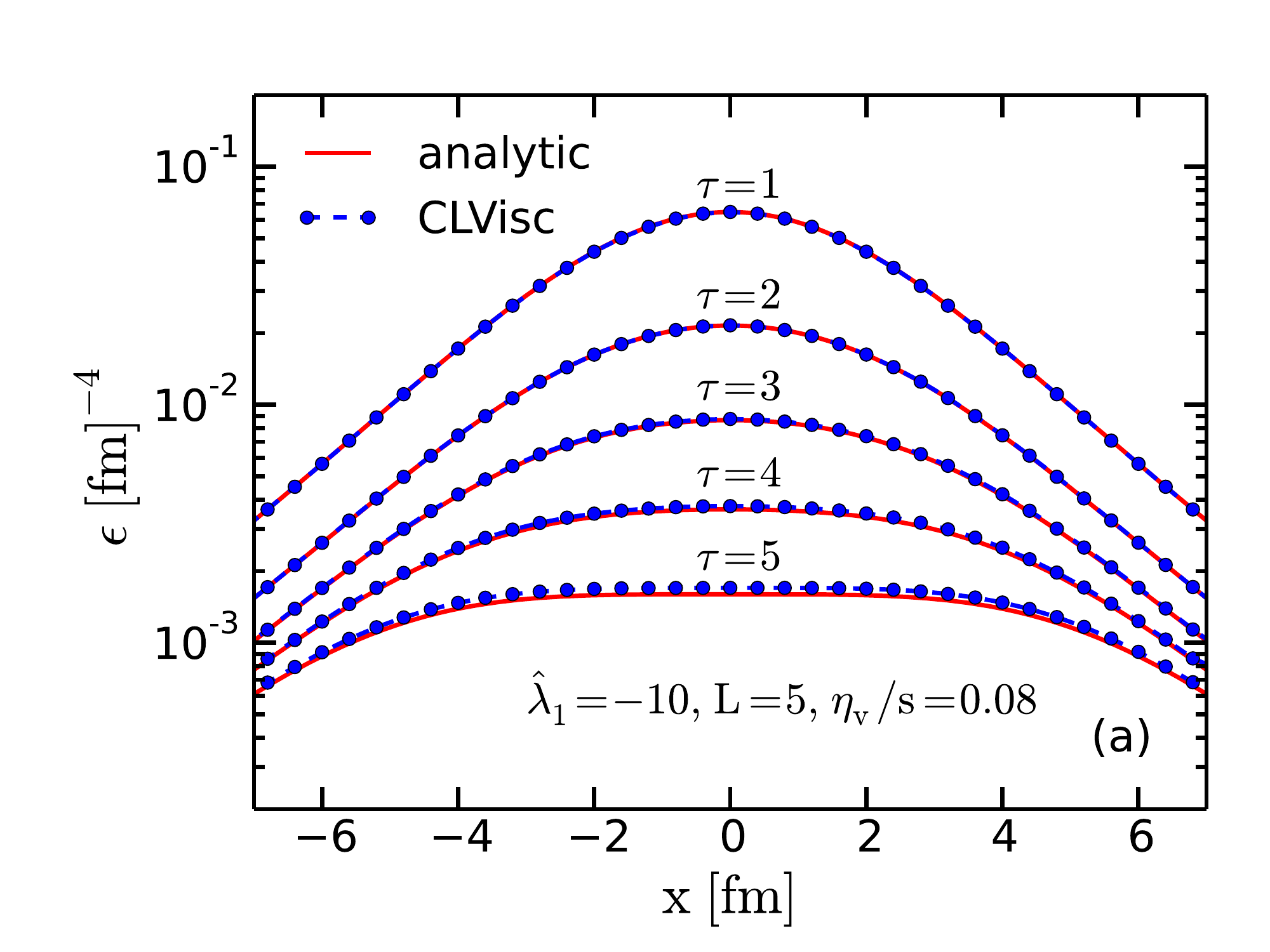}
   \includegraphics[width=70mm]{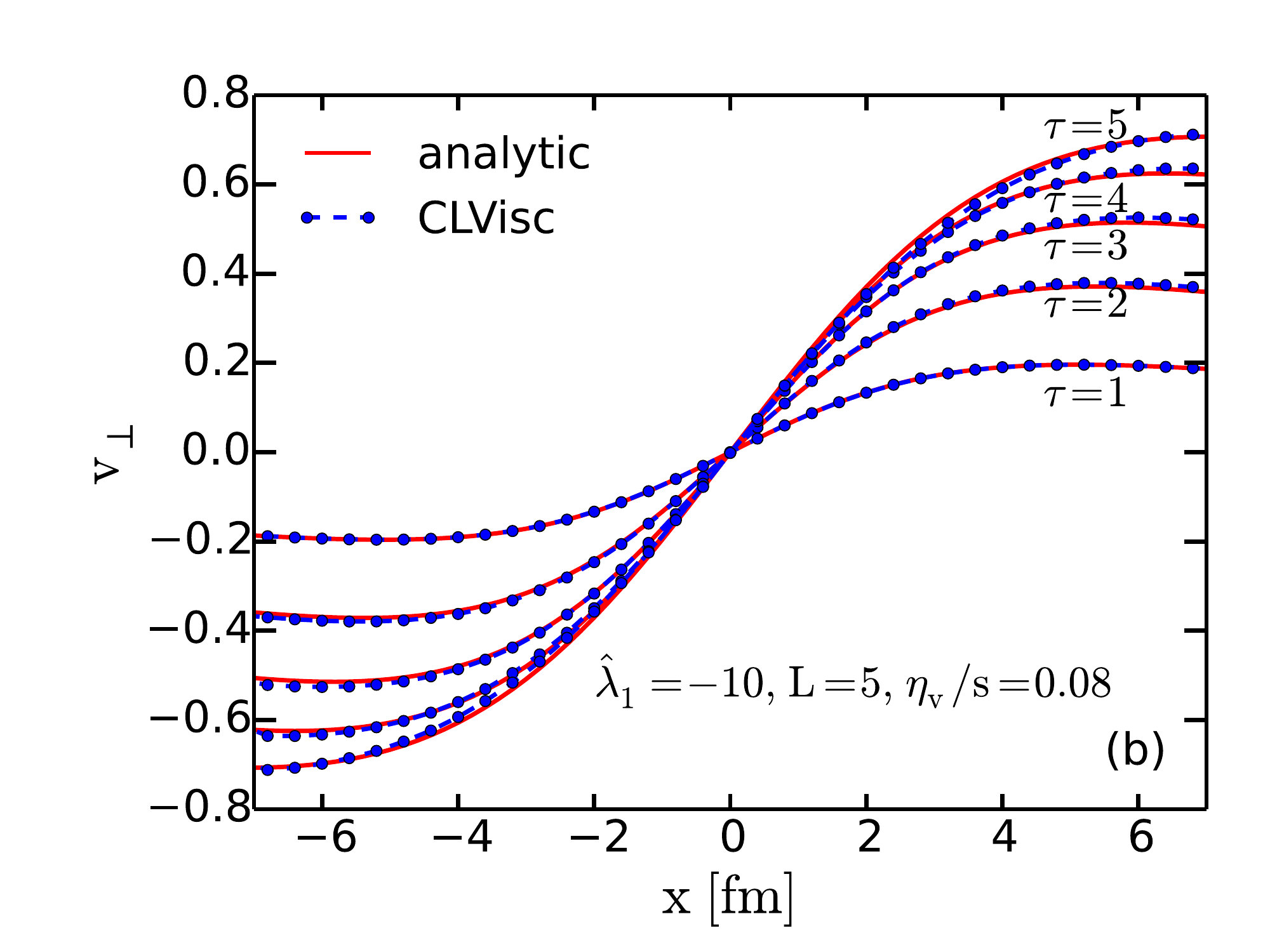}
 \caption{Comparisons of numerical and analytical second-order solutions. Left:  energy density. Right: transverse flow velocity. Figures from \cite{Pang:2014ipa}.}
 \label{fig1}
\end{figure}

\subsection{Analytical vs. numerical second-order solutions}

As mentioned in the introduction, analytic solutions are useful in testing numerical codes.  It is interesting to do this test at the level of second-order hydrodynamics. 
In  \cite{Pang:2014ipa}, an exact second-order solution which generalizes the Gubser flow was found. This is another example of solutions with $\sigma^{\mu\nu}\neq 0$, but it is valid only when the (dimensionless) transport coefficients satisfy a special relation $\eta \lambda_1^2=3\tau_\pi$. Such a relation does not hold in general, but this does not matter since the purpose of \cite{Pang:2014ipa} was to test the accuracy of CLVisc, a new second-order viscous hydro code developed in CCNU-LBNL.  The agreement between the analytical and numerical solutions is perfect, see  Fig.~\ref{fig1}.  Similar tests of analytic solutions have been done in the context of the Israel-Stewart equation \cite{Marrochio:2013wla}, the Boltzmann equation \cite{Denicol:2014xca} and anisotropic hydrodynamics \cite{Nopoush:2014qba}.

\section{Computing flow harmonics $v_n$ analytically}

\subsection{$v_n$ as a function of $n,p_T,\eta, \mu$}  

Now I come to the second main topic. The idea is to compute flow harmonics $v_n$ analytically using the anisotropically deformed Gubser flow. (See \cite{Csanad:2003qa,Csanad:2014dpa} for different approaches.) Consider the $\cos n\phi$-modulation of the Gubser flow as small perturbations  \cite{Gubser:2010ui}\beq
\varepsilon \to \varepsilon (1+\epsilon_n A \cos n\phi)\,,  \nonumber\\
u_\perp \to u_\perp + \epsilon_n B \cos n\phi\,,
\eeq
where $\epsilon_n$ is the eccentricity. 
  The linearized equations for $A$ and $B$, etc. are quite complicated in the viscous case, but one can explicitly solve it in the early time regime $\tau\ll L$ with the result \cite{Hatta:2014jva}
  \beq
\varepsilon =T^4
&\approx& \frac{C^4}{\tau^{4/3}}\frac{(2L)^{8/3}}{(L^2+x_\perp^2)^{8/3}} \left(1-\frac{2\eta/s}{3C}\left(\frac{L^2+x_\perp^2}{2L\tau}\right)^{2/3} \right)^4 
\nonumber \\ && \qquad \quad  \times \left[1-4\epsilon_n \left(1+\frac{2\eta/s}{3C}\left(\frac{L^2+x_\perp^2}{2L\tau}\right)^{2/3} \right) \left( \frac{2Lx_\perp}{L^2+x_\perp^2}\right)^n \cos n\phi \right]\,,
\label{new}r \\ 
u_\perp &=& \frac{2\tau x_\perp}{L^2+x_\perp^2} + \epsilon_n \frac{3nL\tau }{L^2+x_\perp^2} \left( \frac{2Lx_\perp}{L^2+x_\perp^2}\right)^{n-1}
\frac{L^2-x_\perp^2}{L^2+x_\perp^2} \cos n\phi\,,\nonumber \\
u_\phi &=& -\epsilon_n\frac{3n\tau }{2}  \left( \frac{2Lx_\perp}{L^2+x_\perp^2}\right)^n  \sin n\phi\,. \nonumber
\eeq
  Next I plug (\ref{new}) into the Cooper-Frye formula
\beq
(2\pi)^3 \frac{dN}{dY p_T dp_T d\phi_p} &=& \int_\Sigma (-p^\mu d\sigma_\mu)  \left(\exp\left(\frac{u \cdot p + k \mu}{T}\right) + \delta f \right) \nonumber \\
 &\propto& 1+2v_n(p_T)\cos n\phi_p\,, \label{cooper}
\eeq
 where $k=\pm 1$ and $\mu$ is the chemical potential. 
The integral is over the three-dimensional freezeout hypersurface $\Sigma$ which I take to be the surface of constant energy density. Eq.~(\ref{new}) then determines the freezeout time $\tau=\tau_f$ at each position in space 
\beq
\tau_f(x_\perp,\phi) &\approx& \frac{(2L)^5}{B^3(L^2+x_\perp^2)^2}\left(1-\frac{3K  (L^2+x_\perp^2)^2}{2(2L)^4} -3\epsilon_n \left( \frac{2Lx_\perp}{L^2+x_\perp^2}\right)^n    \cos n\phi   \right)\,,
 \eeq
 where $K\propto \eta/s$ is the Knudsen number. 
$B^3$ is a parameter which controls the freezeout time. In order to be consistent with the early time approximation used in (\ref{new}), $B^3$ should be larger than unity $B^3\gg 1$. I then evaluate the three-dimensional integral (\ref{cooper}) analytically to leading order in $1/B^3$ and extract $v_n(p_T)$. Explicit results are available in the low- and high-$p_T$ regions  \cite{Hatta:2014jva}
\beq
\frac{v_n(p_T)}{\epsilon_n}&\approx& \frac{27}{32}\frac{n(n-1)\Gamma(3n)}{\Gamma(4n)}\frac{TK_0(p_T/T)}{p_T K_1(p_T/T)}\left(\frac{64p_T}{B^3T}\right)^n \,, \qquad (p_T \ll B^3T)
\nonumber \\ 
\frac{v_n(p_T)}{\epsilon_n}&\approx& \frac{500p_T}{27TB^3}\left(\frac{\sqrt{5}}{3}\right)^{n-1}\left(n-1-\frac{27K}{200}n\right)\,. \qquad (p_T\gg B^3 T)
\eeq
  At low-$p_T$, $v_n(p_T)\propto p_T^n$, while at high-$p_T$, $v_n(p_T)$ is linear in $p_T$. These are actually generic features of the Cooper-Frye formula, and are not specific to the assumptions of early freezeout or conformal symmetry.
   $p_T$-integrated $v_n$ can also be obtained both at zero \cite{Hatta:2014jva} and finite density \cite{Hatta:2015era} 
  \beq
 \frac{v_n}{\epsilon_n} &=&  \frac{9}{64}\frac{\Gamma(3n)}{\Gamma(4n)}\left(\frac{128}{B^3}\right)^n \Gamma^2\left(\frac{n}{2}\right) \frac{n^2(3n+2)^2(n-1)}{2(4n+1)} \nonumber \\  
&&+ \frac{K}{256}\frac{\Gamma(3n)}{\Gamma(4n)} \left(\frac{128}{B^3}\right)^n \Gamma^2\left(\frac{n}{2}\right)\frac{n^3(n-1)}{3n-1}
 \Biggl\{-\frac{27}{4}(3n^2+3n+2) +9\gamma\left(\frac{3n}{2}+1\right)\left(k-\frac{3f'}{4f}\right) \Biggr\}\,,
\label{final}
\eeq 
where $f(\mu/T)\equiv\varepsilon/T^4$ is a model-dependent function.
 The first line is the ideal case. It contains only one parameter $B^3$ which is  related the freezeout temperature and chemical potential. The second line, proportional to the Knudsen number $K\propto \eta/s$, is the viscous contribution.  The last part proportional to $\gamma$ is the new contribution induced by the finite-density effect (roughly, $\gamma\propto \mu$)  \cite{Hatta:2015era}.  Importantly, this part depends on the charge $k=\pm 1$. I have more to say about this below.

 It immediately follows from (\ref{final}) that $v_n$ decays exponentially at large-$n$
 \beq
 v_n \sim e^{-n\ln (4B^3/27)}\,.
 \eeq
 Quite generally, $B^3$ is an increasing function of density, or equivalently a decreasing function of beam energy. This suggests that it is challenging to experimentally measure higher harmonics in low-energy experiments. It also follows from (\ref{final}) that the viscosity effect grows linearly in $n$
 \beq
 v_n/v_n^{ideal} = 1-{\mathcal O}\left(n \frac{\eta}{s}\right)\,.
 \eeq
 This result disagrees with earlier suggestions in the literature. For instance, Ref.~\cite{Staig:2011wj} proposed a Gaussian behavior $v_n/v_n^{ideal} \sim e^{-n^2 \eta/s}$. 

\begin{figure}[htbp]
 \begin{center}
   \includegraphics[width=70mm]{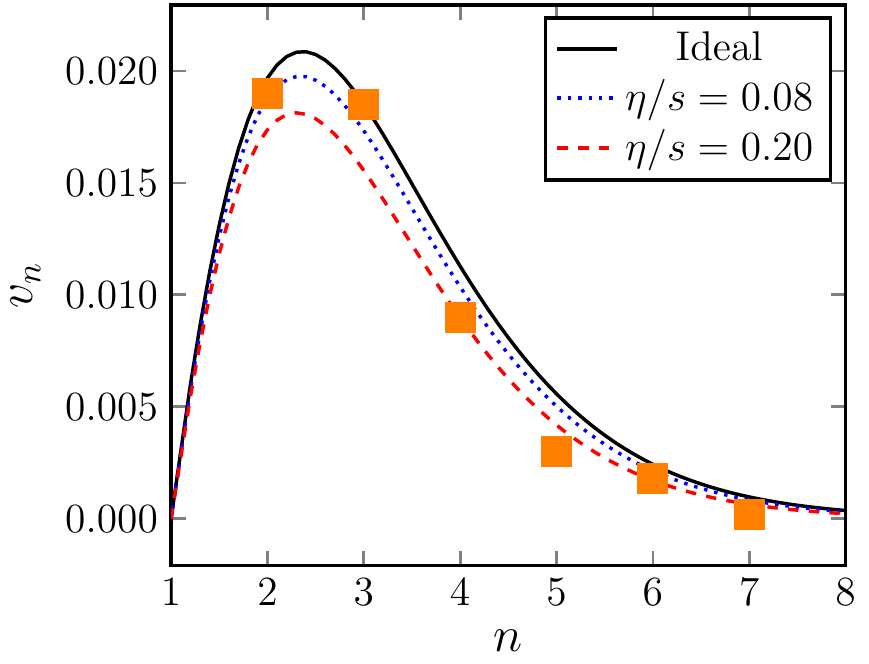}
     \includegraphics[width=65mm]{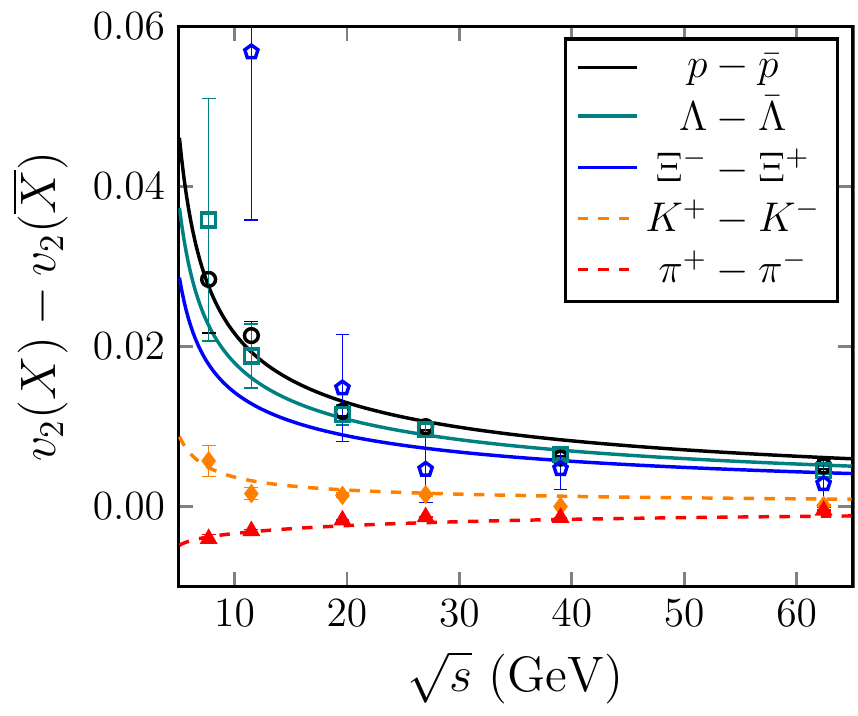}
        \end{center}
 \caption{Left: A plot of (\ref{final}) together with the experimental data from the CMS collaboration for ultra-central collisions. In this plot, all the $\epsilon_n$ are set equal. Right: A fit of the STAR data for $\Delta v_2^X$ using the formula (\ref{from}). Figures from \cite{Hatta:2015era}. }
 \label{fig2}
\end{figure}

 To get an idea of whether (\ref{final}) is reasonable or not,  I plot the function in Fig.~\ref{fig2} (left) together with the CMS data for ultra-central collisions at the LHC \cite{CMS:2012tba}.  In spite of the many assumptions involved in the derivation, the formula (\ref{final}) gives a decent fit, though of course the quantitative agreement should not be taken too seriously. 
 
 \subsection{Charge dependence of $v_n$}
 
Eq.~(\ref{final}) shows that the difference in $v_n$ between particles ($k=1$) and antiparticles ($k=-1$) is proportional to {\it both} the chemical potential and viscosity. The reason of this is simple.   In the Cooper-Frye formula, the charge dependence comes from the fugacity factor
$e^{k\frac{\mu}{T}}$. 
 In ideal hydro, $v_n$ does not depend on $k$ because $\mu/T$ is a constant and drops out in the calculation. 
 However, once viscosity is turned on, $\mu/T$ is no longer a constant but  depends on spacetime coordinates
 \beq
 \frac{\mu}{T}=const. + \frac{\eta}{s} h(x^\mu)\,.
 \eeq
The function $h(x^\mu)$ is integrated over the freezeout surface in the Cooper-Frye formula, and this is how the $k$-dependent term in (\ref{final}) appears. Now consider 
the difference 
\beq
\Delta v^X_2 =v_2^X - v_2^{\bar{X}}\,, \label{sign}
\eeq
for a hadron species $X$ with the quantum numbers $B$, $S$, $I$ (baryon number, strangeness, isospin). Then the following `master formula' can be obtained  from (\ref{final})  \cite{Hatta:2015era}
\beq
\frac{\Delta v^X_2}{v_2^{X,ideal}} \approx  \frac{27 K}{80} \left(\mu_B B+\mu_S S+\mu_I I \right)\,. \label{from} 
\eeq 
In Fig.~\ref{fig2} (right), this formula is used to fit the STAR data for different hadrons species \cite{Adamczyk:2013gw}. 

\subsection{Charged pion $v_2$ difference at RHIC}

Finally, I comment on the recent STAR publication about the charged pion $v_2$ difference \cite{Adamczyk:2015eqo}. The STAR collaboration measured $\Delta v_2^{\pi}$ as a function of the charge asymmetry $A_{ch}=\frac{N_+ -N_-}{N_+ + N_-}$ and found the linear dependence
\beq
\Delta v_2^{\pi^-} \equiv v_2^{\pi^-} - v_2^{\pi^+} = r A_{ch}\,. \label{lin}
\eeq 
The slope $r$ is positive, and has a characteristic peak as a function of centrality. This quantity is of interest because it has been proposed as a signal of the so-called chiral magnetic wave \cite{Burnier:2011bf}. 

Could the STAR result be explained within `normal' hydrodynamics? The charged pions have isospin $I=\pm 1$, so there is some similarity between (\ref{lin}) and  (\ref{from}). However, there is an important difference, the  {\it sign}. In \cite{Hatta:2015era} and above,   $\Delta v_2^X$    is defined as the difference in $v_2$ between positively charged and negatively charged hadrons. In order to comply with the STAR convention (\ref{lin}), I have to reverse the sign in (\ref{from}). Naively,  $\mu_I$ and $A_{ch}$ are roughly proportional to each other with a positive proportionality constant. Then there seems to be a sign mismatch because $r>0$.

 Is it possible to reverse the sign? Here's an argument of how this might happen \cite{Hatta:2015hca}. 
Let me first state in what sense the proportionality $A_{ch}\propto \mu_I$ makes sense. This is not to say that `$\mu_I$ fluctuates event-by-event'. The STAR collaboration measured $\Delta v_2$ in each bin of events labeled by the value of $A_{ch}$ and also centrality. I assign  effective chemical potentials $\mu_{B,S,I}$ in each bin to describe the average properties of the events in this bin.  In the STAR measurements, there are typically ${\mathcal O}(10^5)$ events in each bin. Considering this as a statistical ensemble and assigning $\mu$'s may be marginally justified.     

\begin{figure}[htbp]
   \includegraphics[width=70mm]{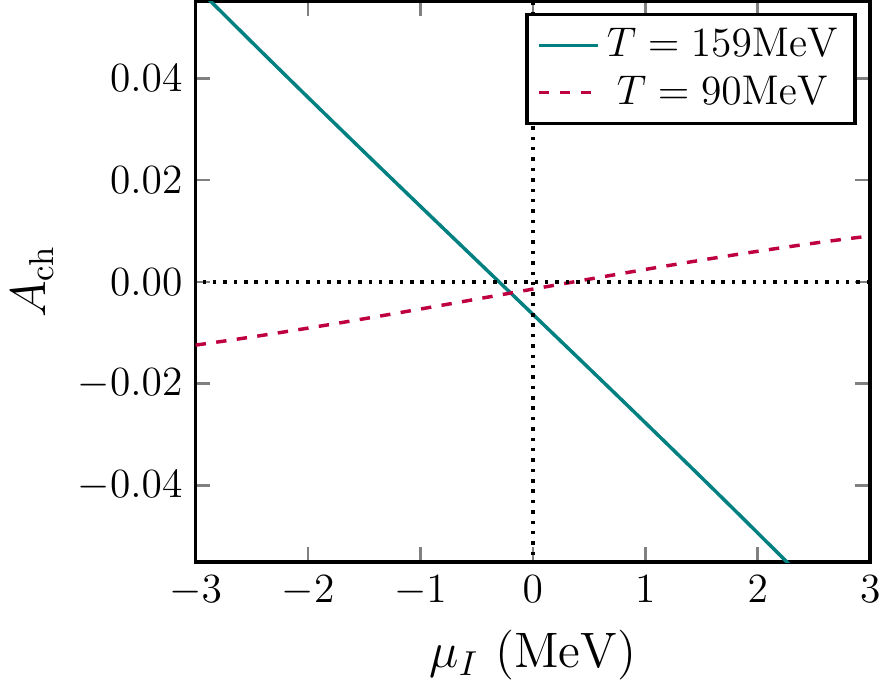}
   \includegraphics[width=70mm]{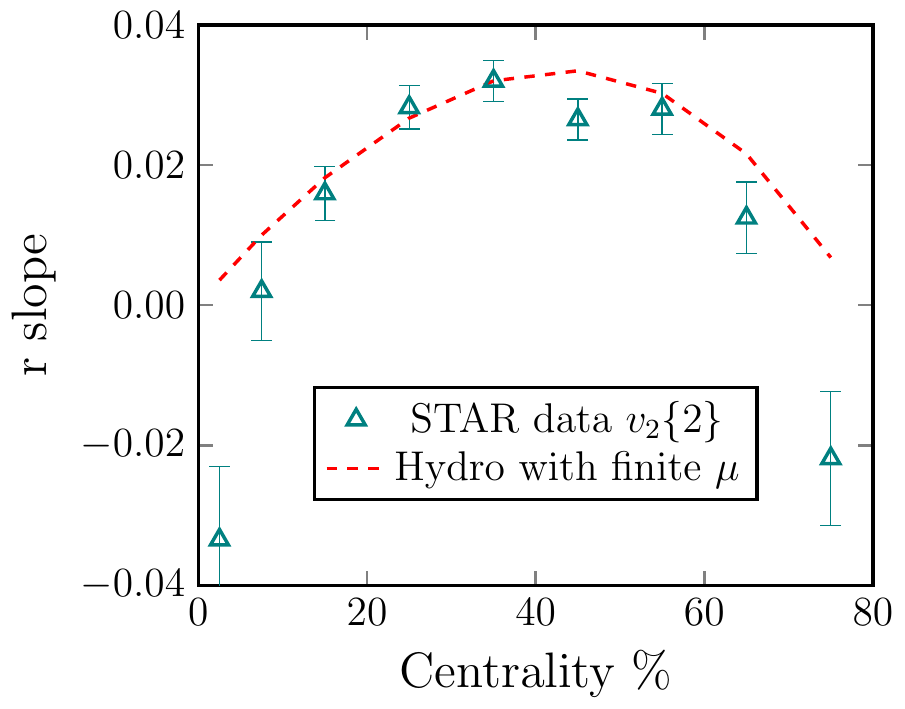}
 \caption{Left: $A_{ch}$ as a function of $\mu_I$ at two different values of $T$. Right: the centrality dependence of the slope $r$.  Figures adapted from \cite{Hatta:2015hca}. }
 \label{fig3}
\end{figure}

Once a set of chemical potentials $\mu_B,\mu_S,\mu_I$ and the temperature is assigned in each bin, the relation between $A_{ch}$ and $\mu$'s can be calculated in the resonance gas model. It is to a good approximation linear  
\beq
A_{ch}=c(T)\,\mu_B + c'(T)\,\mu_I + c''(T)\,\mu_S\,. \label{as}
\eeq
The temperature-dependent coefficients $c,c',c''$ are all positive. This is intuitively clear. The larger $\mu_I$ is, the more $\pi^{+}$ there are in the system, hence larger  $A_{ch}$. However, there are subtleties here. If the freezeout temperature is high, as in the early freezeout scenario which led to the formula (\ref{final}), primordially $A_{ch}$ is not necessarily dominated by $\pi^{\pm}$ since there are many charged hadron resonances. Moreover, in heavy-ion collisions $\mu_B,\mu_I,\mu_S$  are not completely independent of each other. Statistical model fits of hadron yields always find that there is a negative correlation between $\mu_B$ and $\mu_I$
\beq
\mu_I \approx -0.03\mu_B\,,
\eeq
with the coefficient more or less independent of the collision energy. If (and this is a big if) I assume this relation in (\ref{as}), and if the freezeout temperature is not too low, the sign between $A_{ch}$ and $\mu_I$ is reversed, see Fig.~\ref{fig3} (left).  
 
 Now that I get the sign right, I can adjust the value of the Knudsen number $K$ to fit the slope $r$ in one centrality bin. Then the behavior of $r$ in other centrality bins is a prediction because the centrality dependence of the combination $Kv_2^{ideal}$ in (\ref{from}) follows immediately from the results of \cite{Hatta:2015era}. This is in qualitative agreement with the STAR data as shown in Fig.~\ref{fig3} (right).

As a matter of fact, the above scenario faces a severe difficulty in the kaon sector. 
Since $\mu_S>0$, the sign change does not occur and I have to conclude that the slope $r$ for the kaons is negative. Admittedly, this is in conflict with the preliminary STAR data \cite{Shou:2014cja} which suggest that the slope for kaons is also positive. If confirmed with more statistics, I think this is a very nontrivial result, very hard to explain within the conventional  hydrodynamics. 

 \section*{Acknowledgements}

I would like to thank the organizers of Quark Matter 2015 for this opportunity. The results presented here have been obtained in collaborations with Bo-Wen Xiao, Jorge Noronha, Giorgio Torrieri, Long-Gang Pang, Xin-Nian Wang, Mauricio Martinez and Akihiko Monnai, to whom I am grateful.
 




\end{document}